\input{aipcheck}

\documentclass[
    ,final            
  ]
  {aipproc}

\layoutstyle{6x9}

\begin{document}

\title{Observational difference between gamma and X-ray properties
of optically dark and bright GRBs}


\author{L.G. Bal\'azs }{
  address={Konkoly Observatory, Budapest, Hungary}
}

\author{I. Horv\'ath}{
  address={Bolyai Military University, Budapest, Hungary}
}

\author{Zs. Bagoly}{
  address={E\"otv\"os University, Budapest, Hungary}}

\author{A. M\'esz\'aros}{
  address={Charles University, Prague, Czech Republic}}

\author{P. Veres}{
  address={E\"otv\"os University, Budapest, Hungary}}

\begin{abstract}
 Using the discriminant analysis of the multivariate statistical
 analysis we compared the distribution of the physical quantities
 of the optically dark and bright GRBs, detected by the BAT and XRT
 on board of the Swift Satellite.  We found that the GRBs having
 detected optical transients (OT) have systematically higher
 peak fluxes and lower HI column  densities than those without OT.
\end{abstract}

\maketitle


\section{Introduction}

  One of the major tasks of the Swift satellite is to detect and follow up the
  Optical Transients (OTs) accompanying the outburst in the gamma energy range.
  Although, the burst alert  given by the BAT on board of the satellite is
  followed by the XRT detection in the X-ray regime, except a few cases, a
  significant fraction of GRBs remain without detected OT.   Since the vast
  majority of the GRBs are detected  both by  BAT and XRT, it is a
  reasonable question whether there are measurable differences in the
  gamma and /or X-ray properties between the bursts observed by the Swift
  satellite but have or do not have OT.

At its URL location\footnote{
\url{http://swift.gsfc.nasa.gov/docs/swift/archive/grb_table}}
 Swift listed 276 GRB detections until writing the
paper (Oct 26, 2007). Out of these XRT detected 231 and UVOT 72
cases.

In this work we use the $\gamma$ and X-ray data measured by BAT
and XRT: T90 Duration, Fluence, 1-sec Peak Photon Flux, Early
Flux, Initial Temporal Decay Index, Spectral Index   and  Column
Density (NH). We formed two groups from these cases, with and
without OT, and compared them by making use the discriminant
analysis of the multivariate statistical analysis.

\section{Mathematical Summary}

Discriminant analysis aims to make difference between groups in
the multivariate parameter space, orders membership probabilities
to the cases and one may use this scheme for classifying
additional ones not having assigned group memberships. We use this
technique to look for differences in the distributions of GRBs
with or without OT, in the parameter space defined by the BAT and
XRT variables mentioned above.

Let us have a set of $p$ measured variables  on $n$ cases which
are assigned to one of the $k$ classes ($k=2$ in our case). We
look for linear combination of the $x_1,x_2, \ldots,x_p$ variables
which gives maximal separation between the groups of the cases.
There are altogether $k-1$ discriminating variables. In our cases
we have only two groups so we have only one such a variable. It
means we are looking for the variable

\begin{equation}
y=n_1x_1+n_2x_2+ \ldots +n_px_p \phantom{@@@@} where
\phantom{@@@@} \sum^p_{j=1}n^2_j=1
\end{equation}

\noindent with a suitable chosen   $n_1, n_2, \ldots ,n_p$
coefficients ensuring a maximal separation between the two
classes. There are several approaches to solve this problem (for
more details see  \cite{KS:1973}). These are usually among the
major ingredients of the professional statistical software
packages.  We used SPSS\footnote{SPSS is a registered trademark
(\url{http://www.spss.com})} in our computations.

\section{Descriptive Statistics of the Data}

We excluded those cases from our analysis when the slewing time
was greater than 300 sec. With this choice we excluded the GRBs
when the normal slewing of the satellite was blocked by some
reason. Table \ref{tab:a} summarizes the means and standard
deviations of the variables in the analysis for the whole sample
and for the groups, separately.

In the table we listed all cases having measured values in all
variables used in the analysis. We marked with bold face the peak
flux and Hydrogen column density, where the mean values differ
significantly between the groups with and without OT. We give the
results of the test of significance in Table \ref{tab:b}.


\begin{figure}
  \includegraphics[height=.3\textheight]{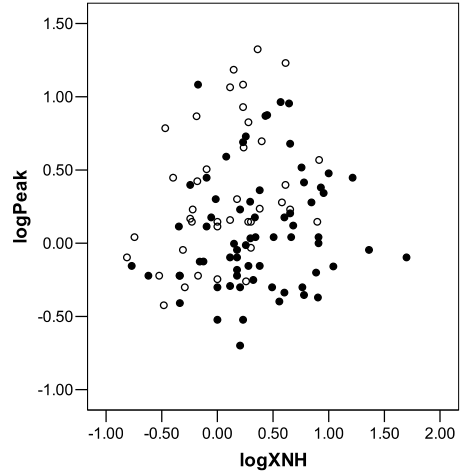}
  \caption{Scatter plot  between the  logarithmic Hydrogen
  column density and  $\gamma$ peak flux. Full dots refer to
  the bursts without and open circles with OT detected by UVOT.
  Note that bursts without OT have significantly higher column
  density and lower peak flux.
}
\end{figure}


\begin{table}
\begin{tabular}{llrrr}
\hline
  \tablehead{1}{r}{b}{OT}
  & \tablehead{1}{r}{b}{Variable}
  & \tablehead{1}{r}{b}{Mean}
  & \tablehead{1}{r}{b}{Standard dev.}
  & \tablehead{1}{r}{b}{No of cas.}   \\
\hline No  &  logT90  &  1.43  &  .67  &  71 \\
   &  logFluence  &  .98  &  .63  &  71 \\
   &  {\bf logPeak}  & {\bf .09}  &  {\bf .42}  &  {\bf 71} \\
   &  logXflux  &  1.18  &  1.31  &  71 \\
   &  logXdec.ind.  &  .18  &  .37  &  71 \\
   &  logXsp.ind.  &  .30 &  .15 &  71 \\
   &  {\bf logXNH}  &  {\bf .35}  &  {\bf .46} &  {\bf 71} \\ \hline
Yes  &  logT90   &  1.43  &  .64  &  33 \\
   &  logFluence  &  1.12  &  .60  &  33 \\
 &  {\bf logPeak}  & {\bf .28}  &  {\bf .45}  &  {\bf 33} \\
   &  logXflux  &  1.26  &  1.13 &  33 \\
   &  logXdec.ind.  &  .14  &  .33  &  33 \\
   &  logXsp.ind.  &  .28  &  .13  &  33 \\
   &  {\bf logXNH}  &  {\bf .12}  &  {\bf .47} &  {\bf 33} \\\hline
Total  &  logT90  &  1.43  &  .66  &  104 \\
   &  logFluence  &  1.02  &  .62  &  104 \\
   &  logPeak  &  .15  &  .43  &  104 \\
   &  logXflux  &  1.20  &  1.25  &  104 \\
   &  logXdec.ind  &  .17&  .36  &  104 \\
   &  logXsp.ind.  &  .30 &  .14  &  104 \\
   &  logXNH  &  .11  &  .46  &  104\\
\hline
\end{tabular}
\caption{Means and standard deviations of the variables in the
analysis for the whole sample and for the groups, separately (we
marked with bold face where the mean values differ significantly
between the groups with and without OT). } \label{tab:a}
\end{table}

\section{Test of significance}

In Table \ref{tab:b} we compared the means of the variables in the
analysis within the groups. We marked with bold face the peak flux
and Hydrogen column density, where the differences in the group
means are significant. $F$ is the test variable denoting the ratio
of the variances between and within the groups.

\begin{table}
\begin{tabular}{lrr}
\hline
  \tablehead{1}{r}{b}{Variable}
  & \tablehead{1}{r}{b}{F value}
  & \tablehead{1}{r}{b}{Significance} \\
  \hline
logT90  &  .000  &  .987 \\
logFluence  &  1.057  &  .306 \\
{\bf logPeak}  & {\bf 4.616}  &  {\bf .034} \\
logXflux  & .103  &  .749 \\
logXdec.ind.  & .248  & .619 \\
logXsp.ind.  & .702  &  .404 \\
{\bf logXNH}  & {\bf 5.499}  &  {\bf .021} \\
\hline
\end{tabular}
\caption{Test of significance of the difference in the mean values
between the groups.} \label{tab:b}
\end{table}

In our case we have two classes. The analysis calculated one
variable to discriminate between our two classes. The level of
significance of this separation is given in Table \ref{tab:c}.

\begin{table}
\begin{tabular}{crrcr}
\hline
  \tablehead{1}{r}{b}{Test of Function}
  & \tablehead{1}{r}{b}{Wilks' $\lambda$}
  & \tablehead{1}{r}{b}{Chi-square}
  & \tablehead{1}{r}{b}{degree of freedom}
  & \tablehead{1}{r}{b}{Significance} \\
  \hline
1  & .880  &  12.877 & 2 &  .002 \\
\hline
\end{tabular}
\caption{Significance of the difference measured by the
discriminant function.} \label{tab:c}
\end{table}

\section{Conclusions}

We performed discriminant analysis in order to look for physical
differences between GRBs with or without OT, detected by the Swift
satellite. We used the following variables measured by BAT and
XRT: T90 Duration, Fluence, 1-sec Peak Photon Flux, Early Flux,
Initial Temporal Decay Index, Spectral Index   and  Column Density
(NH).

The analysis demonstrated a significant difference between the
groups defined. The difference is significant at the 99.8\% level.
The difference is driven by two variables: peak flux of the gamma
radiation and the Hydrogen column density. This latter quantity is
not directly associated to the burst. It rather relates to the
environment.

Our result may indicate that the phenomenon of the OT is connected
with the interaction of the burst with its environment.


\begin{theacknowledgments}
  We are grateful for the valuable discussions with
Stefan Larsson, Peter M\'esz\'aros, Felix Ryde and G\'abor
Tusn\'ady. This study was supported by the Hungarian OTKA grant
No. T48870 and 75072, by a Research Program MSM0021620860 of the Ministry of
Education of Czech Republic, and by a GAUK grant No. 46307 (A.M.).
\end{theacknowledgments}



\bibliographystyle{aipprocl} 


\IfFileExists{\jobname.bbl}{}
 {\typeout{}
  \typeout{******************************************}
  \typeout{** Please run "bibtex \jobname" to optain}
  \typeout{** the bibliography and then re-run LaTeX}
  \typeout{** twice to fix the references!}
  \typeout{******************************************}
  \typeout{}
 }


\end{document}